\documentclass{aa}  

\usepackage{graphicx}
\usepackage{xcolor}
\usepackage{txfonts}
\begin{document}
\title{Magnetic fields in Bok globules: Multi-wavelength polarimetry as tracer across large spatial scales\thanks{Based on observations made with an ESO telescope at the La Silla Observatory under programme ID 096.C-0115. (PI: G. H.-M. Bertrang)}}
\titlerunning{Magnetic fields in Bok globules across large spatial scales}
\author{S. Jorquera
          \inst{\ref{inst1}}
          \and
          G. H.-M. Bertrang\inst{\ref{inst1},\ref{inst2}}
          }
          
\institute{Universidad de Chile, Departamento de Astronom\'ia, Casilla 36-D, Santiago, Chile\\\email{sebastian.jorquera.c@ing.uchile.cl}\label{inst1} \and Max Planck Institute for Astronomy, K\"onigstuhl 17, 69117 Heidelberg, Germany\\\email{bertrang@mpia.de}\label{inst2}}

\abstract
{The role of magnetic fields in the process of star formation is a matter of continuous debate. Clear observational proof of the general influence of magnetic fields on the early phase of cloud collapse is still pending. First results on Bok globules with simple structures indicate dominant magnetic fields across large spatial scales (Bertrang et al. 2014).}
{The aim of this study is to test the magnetic field influence across Bok globules with more complex density structures.}
{We apply near-infrared polarimetry to trace the magnetic field structure on scales of $10^4-10^5\,$au ($\sim 0.05-0.5\textrm{pc}$) in selected Bok globules. The combination of these measurements with archival data in the optical and sub-mm wavelength range allows us to characterize the magnetic field on scales of $10^3-10^6\,$au ($\sim 0.005-5\textrm{pc}$).}
{We present polarimetric data in the near-infrared wavelength range for the three Bok globules CB34, CB56, and [OMK2002]18, combined with archival polarimetric data in the optical wavelength range for CB34 and CB56, and in the sub-millimeter wavelength range for CB34 and [OMK2002]18. We find a strong polarization signal ($P \ge 2 \%$) in the near-infrared for all three globules. For CB34, we detect a connection between  the structure on scales of $10^4-10^5\,$au ($\sim 0.05-0.5\textrm{pc}$) to $10^5-10^6\,$au ($\sim 0.5-5\textrm{pc}$). For CB56, we trace aligned polarization segments in both the near-infrared and optical data, suggesting a connection of the magnetic field structure across the whole globule. In the case of [OMK2002]18, we find ordered polarization structures on scales of $10^4-10^5\,$au ($\sim 0.05-0.5\textrm{pc}$).}
{We find strongly aligned polarization segments on large scales which indicate dominant magnetic fields across Bok globules with complex density structures. To reconcile our findings in globules, the lowest mass clouds known, and the results on intermediate (e.g., Taurus) and more massive (e.g., Orion) clouds, we postulate a mass dependent role of magnetic fields, whereby magnetic fields appear to be dominant on low and high mass but rather sub-dominant on intermediate mass clouds.}


\maketitle

\section{Introduction}\label{intro}

The significance of magnetic fields on the star formation process is subject of ongoing investigations \citep[e.g.,][]{2007PASJ...59..487K, 2009MmSAI..80...54G, 2010ApJ...723..146F, 2015MNRAS.450.4035F, 2015ASSL..407..459L, 2018ApJ...859..165S}. Recent studies show comparatively long lifetimes of massive (e.g., Orion) clouds \citep[$\sim 10\,$Myr; e.g.,][]{2011ApJ...729..133M, 2015ApJ...806...72M} and suggest "dynamically relevant" magnetic fields \citep[e.g.,][]{2008A&A...486L..13A, 2010ApJ...716..299S, 2011ApJ...742L...9C, 2013ApJ...762..120P, 2016A&A...590A...2S, 2016ApJ...819..139B}. The structure of intermediate mass clouds (e.g., Taurus), showing lifetimes of an order of magnitude shorter \citep[$\sim3\,$Myr; e.g.,][]{2001ApJ...562..852H, 2007RMxAA..43..123B,  2016ARA&A..54..135H}, however, can be at least proximately explained by turbulence \citep[e.g.,][]{2001ApJ...546..980O, 2008AJ....136..404O, 2013MNRAS.436.3247K}. In turn, recent studies on Bok globules, the lowest mass clouds known at this time, suggest that globules appear to be magnetically supported \citep{2007ApJ...665..466S, 2011AJ....142...33A, 2014A&A...569L...1A, 2014A&A...565A..94B,CB34optic}. These studies have been conducted on globules with rather simple density structure. In this study, we aim at testing the magnetic field influence on globules with more complex density structures.

The structure of magnetic fields in the dusty envelopes around young stellar objects can be derived from the polarization of background starlight due to dichroic extinction and thermal emission by dust grains \citep[e.g.,][]{2000prpl.conf..247W, 2008Ap&SS.313...87G}. In the case of polarization due to thermal emission, the polarization direction is perpendicular to the magnetic field lines, as the dust grains get partially aligned to the magnetic field, with their major axes perpendicular to the magnetic field lines, projected onto the plane-of-sky \citep[POS; e.g.,][]{2007AAS...21113807H,2007JQSRT.106..225L, 2017MNRAS.469.2869B}. The light of the background stars that runs through the star-forming region also becomes polarized due to dichroic absorption by the dust grains. In this case, the polarization direction directly traces the magnetic field lines, projected onto the POS \citep{2000prpl.conf..247W}.

The influence of magnetic fields on low-mass star formation is best verified when studied independently of other phenomena such as turbulence or stellar feedback. Bok globules \citep{1947ApJ...105..255B} arise as the ideal candidates to study the magnetic field influence on low-mass star forming regions. Bok globules are small, isolated, relative simple structured molecular clouds \citep[e.g;][]{1991ApJS...75..877C}, and are often associated with low-mass star formation \citep{1988ApJS...68..257C, 1994ApJS...92..145Y, 1997A&A...326..329L}. Bok globules are in particular ideal environments to study the correlation between protostellar collapse, fragmentation, and magnetic fields, as these objects are less affected by large-scale turbulence and other nearby star-forming events.

The classical method to characterize magnetic fields is the linearly polarized dust emission and dichroic absorption. Thermal emission of the dust grains, observable in the submillimeter (sub-mm) wavelength range, traces the polarization of the densest, innermost part of the globule, due to the sensitivity of millimeter/sub-mm telescopes. The outer, less dense parts of the globule are traced by the polarization of background starlight due to dichroic absorption, observable in the near infrared (near-IR)/optical wavelength range. Thus, by applying multi-wavelength polarimetry, it is possible to trace the magnetic field on large scales across the globule.

In the following, we will describe our target selection (Section~\ref{desc}), observations (Section~\ref{obs}), and data reduction (Section~\ref{data}). We analyze the polarization maps in Section~\ref{polmap}. In Section~\ref{mfield}, we discuss the magnetic field, its relation to outflows, as well as the magnetic field influence on star formation as a function of mass before we conclude in Section~\ref{conc}.

\begin{table*}[!ht]
\caption{Polarization standard stars.}             
\label{table:standards_summary}      
\centering          
\begin{tabular}{l l c c c c c c c }
\hline\hline       
Instrument & Object & $\alpha_{2000}$ & $\delta_{2000}$ & Type & P & $\gamma$ & Filter & Ref.\\
& & (hh:mm:ss.ss) & (dd:mm:ss.ss) & & (\%) &  $\left(^\circ\right)$ & & \\
\hline                    
SOFI/NTT  & CMa R1 No.24 & 07:04:47.36 & -10:56:17.44 & polarized   & $2.1\pm0.05$    & $86\pm1$    & J & 1 \\
	     & HD64299      & 07:52:25.51 & -23:17:46.78 & unpolarized & $0.151\pm0.032$ &             & B & 2 \\
	     & WD0310-688   & 03:10:31.02 & -68:36:03.39 & unpolarized & $0.051\pm0.09$  &             & V & 3 \\
\hline                  
\end{tabular}
\tablebib{ (1) \citet{1992ApJ...386..562W}; (2) \citet{1990AJ.....99.1243T}; (3) \citet{2007ASPC..364..503F}.}
\end{table*}

\section{Description of the sources}\label{desc}

Our aim is to verify the large-scale magnetic field influence on low-mass star formation. As stated in Section~\ref{intro}, Bok globules are ideal candidates for this kind of study. Furthermore, we have to restrict us to globules with available archival sub-mm or optical polarization maps that trace the magnetic field structure at scales not observable in the near-IR wavelength range. We further aim at expanding our previous study of the magnetic field structure in low-mass star-forming regions \citep{2014A&A...565A..94B} to more complex structured objects. We found that these criteria are satisfied by three Bok globules: CB34, CB56, and [OMK2002]18.

\textit{CB34} is a Bok globule located at a distance of $\sim 1.5\,$kpc \citep{1997A&A...326..329L}. The inner region of the globule discloses three dense cores (C1, C2, C3), each extends by $\sim 2.5\,$pc in diameter \citep{2003MNRAS.344.1257C, 2010ApJS..188..139L}. Their masses are $52\,\rm{M}_\odot \, (\rm{C1}), \, 94\,\rm{M}_\odot \, (\rm{C2}), \, \rm{and} \, 13\,\rm{M}_\odot \, (\rm{C3})$, respectively \citep[Fig.~\ref{CB34map};][]{2003MNRAS.344.1257C}. These cores exhibit ongoing star formation based on the presence of a water maser as well as several collimated outflows \citep[Fig. 1;][]{2006AJ....132.1322G, 2000A&A...362..635H, 1994ApJS...92..145Y}. The chemical age of the globule is estimated to be >$10^5$ yr \citep{2010ApJS..188..139L}, with the presence of a pre-main-sequence star with an age of $\sim 10^6$ yr located in the core C1 \citep{1997AJ....113.1395A}. This massive globule encloses within its center ($4.05\times2.25)\times10^5\,$au ($\sim 2 \times 1\textrm{pc}$) about $110-170 M_\odot $ \citep{2010ApJS..188..139L,2003MNRAS.344.1257C, 1998ApJS..119...59L}.

\textit{CB56} is a compact, irregular shaped cloud with a major and minor axis of $4\farcs5$ and $2\farcs5$ respectively \citep{2002A&A...383..631D}. Its major axis has a position angle of $170^{\circ}$ with respect to the galactic plane. CB56 is associated with two infrared point sources \citep[IRAS 07125-2503 and IRAS 07125-2507;][]{1988ApJS...68..257C}. The distance to CB56 has yet to be determined.  The optical polarization signal of this globule is significantly stronger than the interstellar polarization in the optical wavelength range within $200\,$pc \citep[][Gontcharov, priv. comm.]{2014MNRAS.442..479C, 2017ASPC..510...78G}. Together with the strong near-IR polarization, this indicates that CB56 is located within a distance of $200\,$pc and the measured polarization signal is associated with CB56.

\textit{[OMK2002]18} is a low-mass star forming region at a distance of $140\,$pc \citep{2011A&A...536A..99M}. It consist of the two dense clouds [OMK2002]18a and [OMK2002]18b. Both clouds have a radius of $\sim4.125\times10^3\,$au and enclose masses of $2.3\,\rm{M}_\odot \, \rm{and} \, 1.1\,\rm{M}_\odot$ respectively \citep{2002ApJ...575..950O}. OMK[2002]18 is associated with one compact infrared source \citep[IRAS 04191+1522;][]{2002ApJ...575..950O}.  
    
\section{Observations}\label{obs}
The observations were performed in the near-IR with the instrument Son OF ISAAC (SOFI) at the  $3.58$~m New Technology Telescope (NTT) from November 11th to 15th, 2015.  SOFI/NTT is mounted at the Nasmyth~A focus, equipped with a $1024\times 1024$ Hawaii Rockwell array optimized for wavelengths of $1-2.5~\mu$m.\\
In SOFI/NTT, a single Wollaston prism is used for polarization observations. In this observing mode, the polarized flux is measured simultaneously at two different angles that differ by $90\degr$. To derive the linear polarization degree and orientation of an object, two observations must be performed at each pointing of the telescope with different orientations of the Wollaston prism, typically $0\degr$ and $45\degr$. This is realized by a rotation of the complete instrument. To avoid overlapping between different polarization images an aperture mask of three alternating opaque and transmitting strips of about $40''\times300''$ for SOFI/NTT is used.\\
We carried out Js-band polarization observations of three fields of CB34 and [OMK2002]18ab as well as one field of CB56.

\section{Data Reduction}\label{data}
For the data reduction, we apply an specialized pipeline created to work with polarization data obtained with SOFI/NTT \citep{2014A&A...565A..94B}. This pipeline performs bias correction, flat-fielding, and instrumental polarization extraction as well as the calculation of the Stokes parameters \textit{I}, \textit{Q}, and \textit{U} via aperture photometry. Table \ref{table:standards_summary} summarizes the information of polarized and unpolarized standard stars used to determine the instrumental polarization.

A correction factor for the Wollaston prism was computed for the reduction of the instrumental polarization (see Appendix). This correction factor was tested for the polarized standard star, and as a successful test, the polarization data obtained after bias correction corresponds well to literature values.

\begin{figure*}
\centering
  \includegraphics[width=17cm]{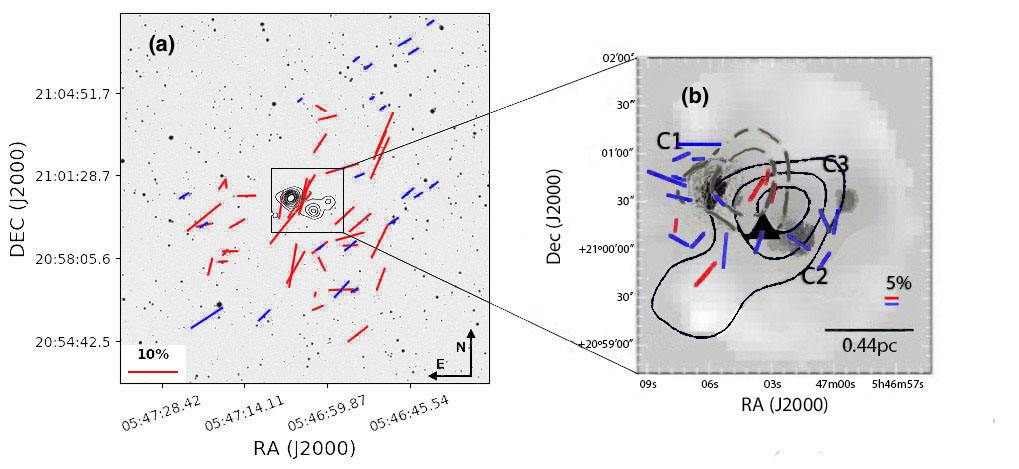}
    \caption{(a) Near-IR polarization segments (red) and optical polarization segments (blue) are plotted over a $15' \times 15'$ R-Band DSS image of the field containing CB34. A 10\% polarization {segment} is drawn as reference for both data sets into the lower left corner. Contours correspond to JCMT/SCUBA 850 $\mu$m dust continuum emissions from 0.1 to 0.7 Jy $\rm{beam}^{-1}$, with intervals of 0.1 Jy $\rm{beam}^{-1}$ \citep[][]{CB34optic}. (b) Zoom into the dense center of CB34. The near-IR polarization segments (red) and the archival sub-mm polarization segments (blue) are plotted over the total continuum intensity map at 850 \citep{CB34optic}. The sub-mm segments are binned over a $10'' \times 10''$ grid. The solid (dashed) black (gray) contour lines correspond to blueshifted (redshifted) integrated CO emission. Contour levels are spaced at 0.1 K km $\rm{s}^{-1}$ intervals of 0.5 K km $\rm{s}^{-1}$  (black), and 0.2 K km $\rm{s}^{-1}$ intervals of 0.8 K km $\rm{s}^{-1}$ (white), respectively (from \citet{1994ApJS...92..145Y}, beam size of 48").  A 5\% polarization {segment} is drawn as reference for both data sets into the lower right corner. The well-ordered polarization pattern indicates dominant magnetic fields on scales of $10^3-10^6\,$ au ($\sim 0.005-5\textrm{pc}$).} 
    \label{CB34map}
\end{figure*}

\begin{figure*}
\centering
  \includegraphics[width=17cm]{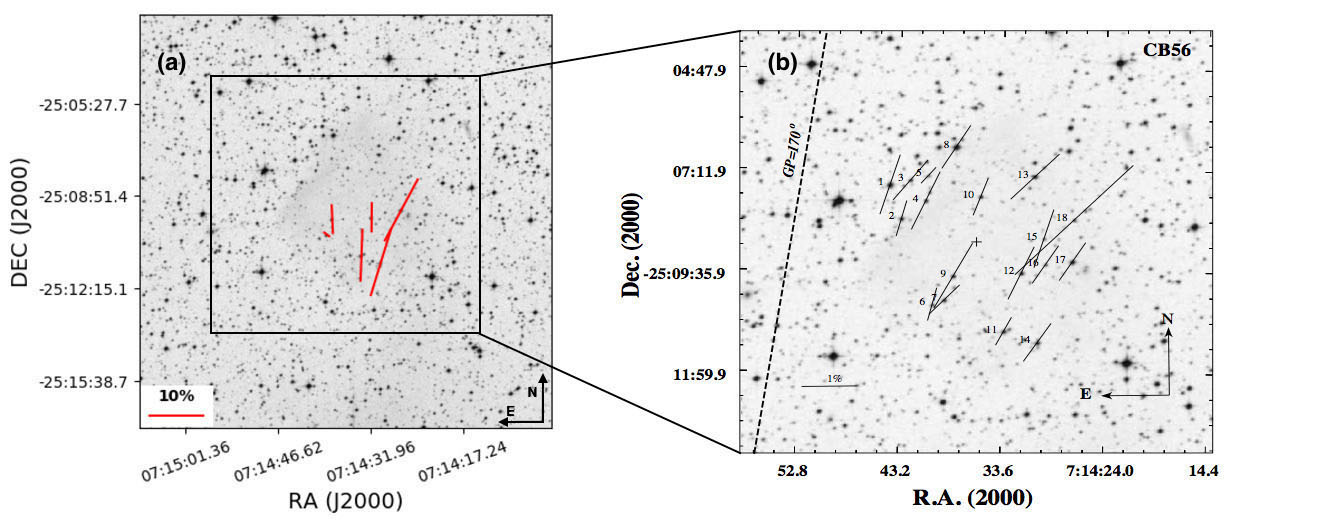}
    \caption{(a) Near-IR polarization segments plotted over a $15' \times 15'$ R-Band DSS image of the field containing CB56. A 10\% polarization {segment} is drawn for reference in the lower left corner. Only the segments with $P/\sigma_p \ge 3$ are plotted. (b) Optical polarization segments of CB56, plotted over a $10' \times 10'$ R-Band DSS image  \citep{2014MNRAS.442..479C}. A 1\% polarization {segment} is drawn for reference in the lower left corner. The near-IR and optical polarization is strong and well-ordered, indicating dominant magnetic fields on scales of $10^4-10^5$ au ($\sim 0.05-0.5\textrm{pc}$).}
    \label{CB56map}
\end{figure*}

\begin{figure*}
\centering
  \includegraphics[width=17cm]{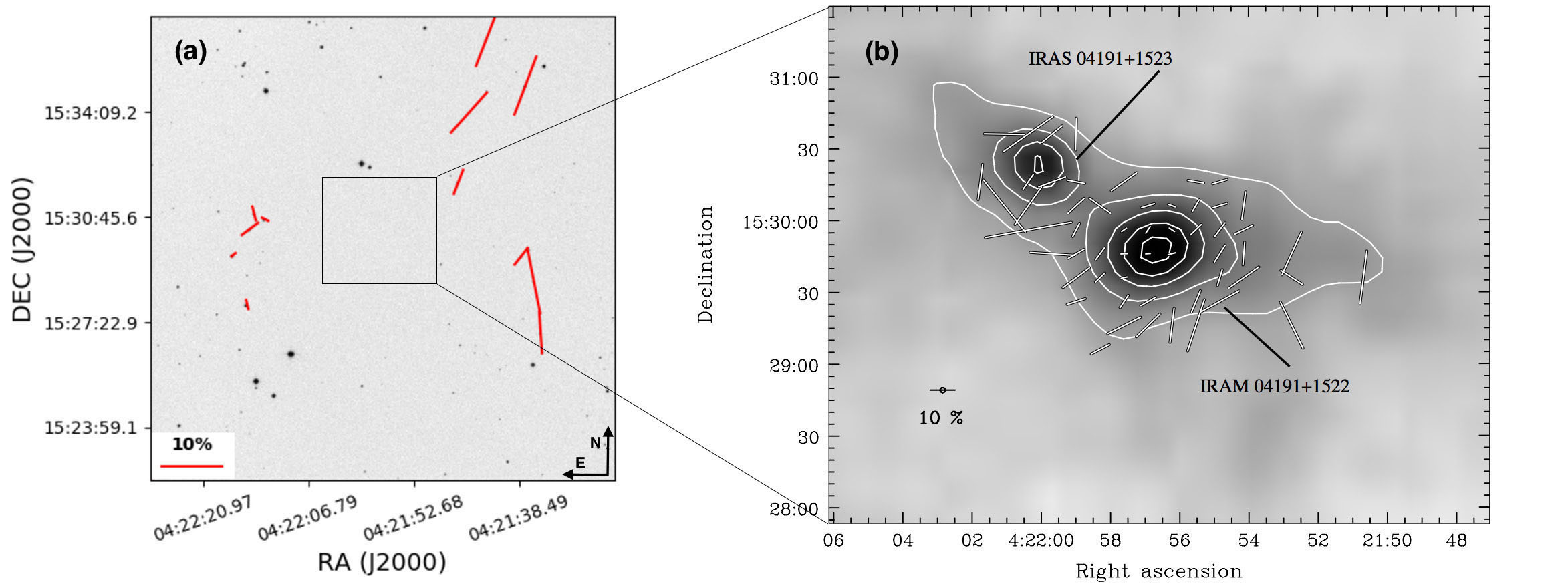}
    \caption{(a) Near-IR polarization segments plotted over a $\sim 20' \times 20'$ R-Band DSS image of the field containing [OMK2002]18. A 10\% polarization {segment} is drawn for reference in the lower left corner. Only the segments with $P/\sigma_p \ge 3$ were plotted. (b) Sub-mm polarization segments of [OMK2002]18, plotted over a $20'' \times 20''$ grid. Contours correspond to JCMT/SCUBA 850 $\mu$m dust continuum emissions from 0.05 to 0.25 Jy $\rm{beam}^{-1}$, with intervals of 0.05 Jy $\rm{beam}^{-1}$ \citep[][\copyright AAS. Reproduced with permission]{2009ApJS..182..143M}. Only segments with $P/\sigma_p \ge 2$ are plotted. The near-IR polarization is strong and well-ordered, indicating dominant magnetic fields on scales of $10^4-10^5$ au ($\sim 0.05-0.5\textrm{pc}$) in the western region of the globule.}
    \label{OMKmap}
\end{figure*}

\begin{figure*}
\centering
  \includegraphics[width=17cm]{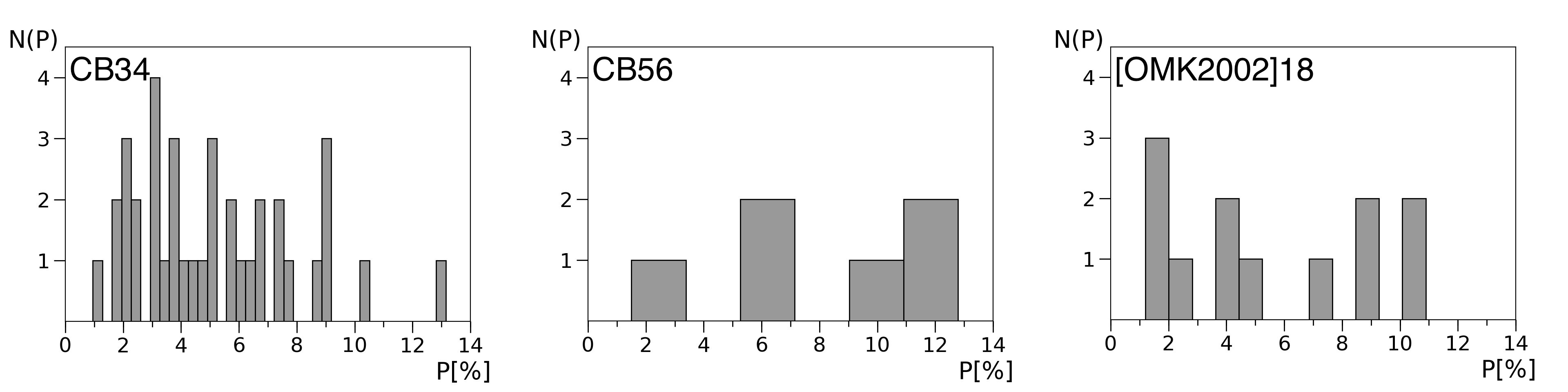}
    \caption{Distribution of polarization degree, P for $P \ge 3 \%$, counts given by $N(P)$, of CB34, CB56, and OMK[2002]18, observed in the near-IR.}
    \label{pol_degrees}
\end{figure*}

\begin{figure*}
\centering
  \includegraphics[width=17cm]{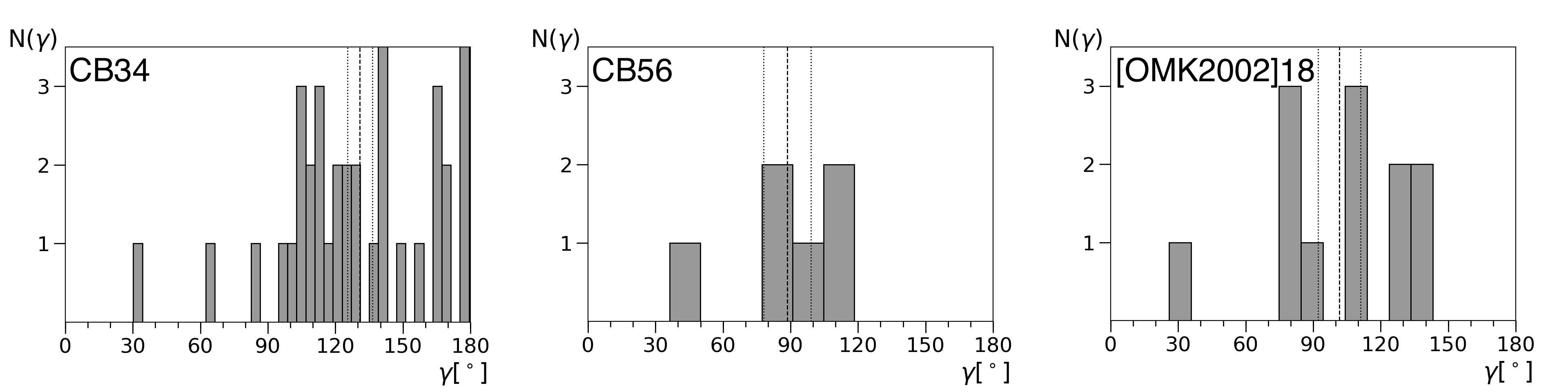}
    \caption{Distribution of polarization angle, $\gamma$, counts given by $N(\gamma)$  of CB34, CB56, and OMK[2002]18, observed in the near-IR. The dashed lines represent the mean polarization angles, while dotted lines represent the corresponding 1$\sigma$ deviation.}
    \label{pol_angles}
\end{figure*}

\section{Polarization Maps}\label{polmap}
In this section, we present the polarization maps of the three Bok globules CB34, CB56, and OMK[2002]18. We compare our near-IR data to archival polarization data in the optical and sub-mm wavelength range. As stated in Section~\ref{intro}, the polarization at these different wavelength ranges arises from different sources: while the polarization in the sub-mm range arises from thermal emission, in the near-IR and the optical it is caused by dichroic absorption. Thus, the polarization of a dust grain observed in the sub-mm is oriented perpendicular to the polarization of the very same dust grain observed in the near-IR or optical. 

For CB34, {in addition to our new data in the near-IR wavelength range, archival data is available in both the optical and sub-mm wavelength ranges \citep{CB34optic}}. This enables an analysis over the whole range of scales in this globule, from its most dense center (sub-mm) towards the less dense outer envelope (near-IR, optical).

For CB56, the data set is restricted to archival data in the optical wavelength range \citep{2012BASI...40..113P,2014MNRAS.442..479C} in addition to the newly obtained near-IR data. However, those two wavelength ranges cover this object almost completely. For this comparison, we make use of the optical polarization data published in \citet{2014MNRAS.442..479C}, given that it is the more comprehensive data set at this wavelength range.

For [OMK2002]18, archival data only in the sub-mm wavelength range were available \citep{2009ApJS..182..143M}, in addition to the newly obtained near-IR polarization data. With these, an analysis of the polarization of the most dense center of the globule is possible (sub-mm), as well as an analysis of the less dense outer regions (near-IR).

In the following, we discuss the polarization maps separately, first focusing on the new near-IR data, and subsequently, comparing it to the maps of the optical and sub-mm wavelength ranges.

\subsection{CB34}
The polarization maps of the Bok globule CB34 are shown in Fig. \ref{CB34map}. CB34 disclose a strong polarization signal, $P\ge1\%$, in the outer, less dense regions observed in the near-IR (Fig. \ref{pol_degrees}, Fig. \ref{pol_angles}). The polarization segments in the near-IR wavelength range cover this Bok globule on scales of $10^4-10^5\,$au ($\sim 0.05-0.5\textrm{pc}$). {The polarization segments in the near-IR wavelength range are well-ordered and oriented towards North-West with a mean orientation of $111^\circ \pm 6^\circ$.} 

Most of the outer regions of this globule are traced in both near-IR and optical polarization. Both data sets fit well in orientation ({within the errors}). The mean value of the polarization angle in the near-IR wavelength range in this regions is relatively consistent with the position angle of the galactic plane at the latitude of CB34, $\theta_{\rm{GP}}\approx 148.7^\circ$, which suggests that the magnetic field is coupled to the Galactic magnetic field in the observed POS. This is similar to the findings in the optical \citep[][]{CB34optic}

A puzzling finding is the relation between the polarization degrees in the near-IR and the optical. As opposed to the empirical Serkowski relation \citep{1975ApJ...196..261S}, the near-IR signal is stronger polarized than the optical. Similar discrepancies have been found also in very few other globules, tentatively explained by either a limited location of origin of polarization or the presence of two different dust populations \citep{2012ApJ...748...18C, 2014A&A...569L...1A}. Recently, it has been suggested that not only the grain size but also the chemical composition of grains alter the extinction-dependency of polarization  \citep{2018MNRAS.tmp.1459P}. Future theoretical work is needed to investigate this in more detail.

In the sub-mm, the polarization segments can be associated to the cores C1 and C2, while there is no detection on top of C3 (Fig. \ref{CB34map}). The region observed in the sub-mm overlaps with our near-IR data. The polarization segments in the sub-mm wavelength range were plotted using the original data from \citet[][]{CB34optic}. Thus, it has to be keep in mind that the segments must be rotated $90^\circ$ to be aligned with the magnetic field orientation. In this region, the orientation of the polarization segments obtained in the near-IR wavelength range fit very well to the sub-mm polarization segments associated with core C1, but deviate stronger from those of core C2. {The polarization segments in the near-IR wavelength range are located just in between the cores C1 and C2}. In this region, the mean polarization angle in the near-IR wavelength is $\bar{\gamma}_{\rm{near-IR}}\approx 111.9^\circ \pm 11.5^\circ$, which suggest that in this case the orientation of the magnetic field in the center of the globule is different to the Galactic magnetic field. Similar findings where obtained in \citet{CB34optic} using the data from the sub-mm wavelength range.

{The change in the polarization pattern from the outer to the central part of CB34 indicates that the magnetic field is coupled on large-scales to the galactic magnetic fields, while it disconnects in the center where the globule is fragmented into several cores. The scale of the transition is $\sim 10^4$ au ($\sim 0.05\textrm{pc}$; Fig. \ref{CB34map}).}

\subsection{CB56}
The polarization maps of the Bok Globule CB56 are shown in Fig. \ref{CB56map}. CB56 discloses a strong polarization signal with polarization degrees of mostly $5\%-12\%$ (Fig.~\ref{pol_degrees}). The polarization structure of this globule in the near-IR is well-ordered and predominantly oriented towards North-North/West (Fig.~\ref{pol_angles}).

The region of CB56 which is covered by our near-IR observations overlays with the archival optical data set \citep[][]{2014MNRAS.442..479C}. {The globule discloses a weak polarization signal in the optical wavelength range, with an average of $P\approx 1\%$(see Section 5.1 for a discussion about the Serkowsky relation).} 

The polarization segments obtained in both the optical and near-IR wavelength ranges are very well aligned within themselves, and fit well to each other. We find a slight deviation of the polarization orientation towards the dense center of the globules when switching from the optical to the near-IR wavelength range.

\subsection{OMK[2002]18}

The polarization maps of OMK[2002]18 is shown in Fig. \ref{OMKmap}.  OMK[2002]18 shows a strong polarization signal, $2 \ge P \ge 10\%$, in all the regions observed in the near-IR (Fig. \ref{pol_degrees}).

Two clearly divided regions of the globule can be observed in the near-IR wavelength range. The western region of the globule shows well-ordered polarization segments in North/North-West orientation with strong polarization degrees of $P \approx 8 \%$. The eastern region of the globule shows an unordered pattern with a much weaker polarization signal, $P \approx 3 \%$ indicative for depolarization effects \citep[e.g., ][]{2016A&A...588A.129B}. Considering the separation between these regions, uniformly ordered polarization segments are not necessarily expected. 

A spatial gap is seen between the sub-mm an near-IR polarization segments, which of about $3.3 \times 10^4$ au ($\sim 0.16\textrm{pc}$). It originates in the sensitivity of SOFI/NTT, resp. SCUBA/JCMT \citep{2014A&A...565A..94B}. The near-IR polarization segments in the western part of this object are well-aligned with the most external polarization segments in the sub-mm wavelength range.

\section{Magnetic Fields}\label{mfield}
In our analysis we assume that the magnetic field is oriented perpendicular to the measured polarization segments in the sub-mm and parallel oriented to the measured polarization segments in the near-IR and in the optical. This widely applied concept is based on the finding that, independently of the alignment mechanism, charged interstellar dust grains would have a substantial magnetic moment leading to a rapid precession of the grain angular momentum around the magnetic field direction, implying a net alignment of the grains with the magnetic field \citep[e.g., ][]{1997ApJ...480..633D, 2007JQSRT.106..225L}. However, one has to keep in mind that polarization observations strongly suffer from projectional effects along the line-of-sight (LOS). Thus, for a comprehensive understanding of the magnetic field structure additional 3D radiative transfer modeling is essential but beyond the scope of this study.

In general, a high polarization signal is connected to a magnetic field strong enough to align the dust grains along the LOS. In Section \ref{polmap}, we find well-ordered polarization segments on scales of $10^3\,$au to $10^5\,$au ($\sim 0.005-0.5\textrm{pc}$) in all of the three Bok globules, CB34, CB56, and OMK[2002]18. This implies the presence of dominant large-scale magnetic fields in each globule.

In addition to this analysis, we can further estimate the magnetic field strength in the POS by using a method first proposed by \citet{1953ApJ...118..113C} (CF method):
\begin{equation}\label{eq:CF}
|B_{\rm{POS}}| = \alpha\cdot\sqrt{\frac{4\pi}{3}\rho_{\rm{gas}}}\frac{\upsilon_{\rm{turb}}}{\sigma_\gamma},
\end{equation}

where $\rho_{\rm{gas}}$ is the gas density in units of g $\rm{cm}^{-3}$, $\upsilon_{\rm{turb}}$ is the rms turbulence velocity in units of cm $\rm{s}^{-1}$, and $\sigma_\gamma$ the standard deviation of the polarization angles in radians. It is assumed that the magnetic field is frozen in the cloud material.

Originally, $\alpha = 1$ for the CF' method. Different studies \citep[e.g,][]{1990ApJ...362..545Z, 1991ApJ...373..509M} have suggested a lower value for $\alpha$. \cite{2001ApJ...546..980O} found that, for $\sigma_\gamma \lesssim 25^\circ (0.44\, \rm{radians})$, using $\alpha \sim 0.5$ yields a good estimate for the magnetic field strength.   

The total gas density, $\rho_{\rm{gas}}$, is given by:
\begin{equation}\label{eq:gasden}
\rho_{\rm{gas}} = 1.36 \,  n_{\rm{H}_2} \, M_{\rm{H}_2},
\end{equation}
where $M_{H_2} = 2.0158$ amu is the mass of a $\rm{H}_2$ molecule.

The rms turbulence velocity, $\upsilon_{\rm{turb}}$, is given by \citep{1995ApJ...454..217W}:
\begin{equation}
\upsilon_{\rm{turb}} = \frac{\bigtriangleup\upsilon}{2.35},
\end{equation}
where $\bigtriangleup\upsilon$ is the FWHM line width, measured at quiescent positions located away from the emission peaks.

For the outer, less dense regions of CB34, the mean density, $n_{\rm{H}_2}$, is $3.76\times10^2 \, \rm{cm^{-3}}$ \citep{CB34optic} on average, while the FWHM line width, $\bigtriangleup\upsilon$, corresponds to 1.37 km $\rm{s}^{-1}$ \citep{2011A&A...527A..41S}. 
Considering the globule's morphology, we split CB34 into two regions: South-East (SE) and North-West (NW) of the center. Our near-IR observations trace the central region itself too sparsely for a robust application of the CF method. All parameters used for each region are listed in Table \ref{table:field_strength}.

We determine the dispersion of the polarization angles in the SE region to $\sigma_{\gamma}=31^\circ$ and in the NW region to $\sigma_{\gamma}=29^\circ$. Since the correction found by \citet{2001ApJ...546..980O} is not applicable, the derived magnetic field strengths are upper limits. Applying Equation~(\ref{eq:CF}) ($\alpha=1$), we derive magnetic field strengths of $B^{\rm{NW}}_{\rm{NIR}} \approx 9.6\,\mu\rm{G}$ (North-West) and $B^{\rm{SE}}_{\rm{NIR}} \approx 9.1\,\mu\rm{G}$ (South-East), respectively. The similar magnetic field strengths in both regions, along with the well-ordered polarization segments (Section \ref{polmap}), implies a uniform large-scale magnetic field across the globule.

Towards the globule's center, the magnetic field strength increases to $34\,\mu\rm{G}$ for core C1 and $ 70\,\mu\rm{G}$ for core C2 \citep{CB34optic}. This is consistent with the theoretical picture of magnetically supported star formation \citep[e.g.,][]{2011ApJ...729...72P, 2009A&A...506L..29H, 2016JPhCS.719a2002F}.

For CB56 and OMK[2002]18, necessary information on the gas densities and turbulence velocities in the regions traced by the presented polarization data is not available.

\begin{table*}[!ht]
\caption{Gas densities, gas velocities, polarization, and magnetic field strengths of the regions of CB34 traced in the near-IR.}             
\label{table:field_strength}      
\centering          
\begin{tabular}{l c c c c c c c c }
\hline\hline       
Region & $n_{\rm{H}_2}$ & $\rho_{\rm{gas}}$ & $\bigtriangleup\upsilon$ & $\upsilon_{\rm{turb}}$ & $N_{\rm{vec}}$ & $\bar{\gamma}$ & $\sigma_\gamma$ & B\\
& $(\rm{cm^{-3}})$ & (g $\rm{cm^{-3}}$) & (km $\rm{s^{-1}}$) & (km $\rm{s^{-1}}$) &  & (rad) &  (rad) & ($\mu\rm{G}$) \\
\hline                    
North-West  & $3.76\times10^2$ & $1.71\times10^{-21}$ & 1.37 & 0.58 & 11 & 2.32 & 0.51 & 9.6 \\
South-East  &  $3.76\times10^2$ & $1.71\times10^{-21}$ & 1.37 & 0.58 & 12 & 2.35 & 0.54 & 9.1 \\
\hline                  
\end{tabular}
\end{table*}

\subsection{Correlation between magnetic field structure and the CO outflow}
Alongside the collapse of the dust in Bok globules, magnetic fields are also presumed to influence the formation of circumstellar disks and outflows \citep[e.g.,][]{2006ApJ...637L.105M, 2017MNRAS.464L..61B, 2017MNRAS.469.2869B}. Both aligned and misaligned orientation of the outflow axes along the magnetic field directions have been reported \citep[e.g.,][]{2003AJ....125.1418J, 2014A&A...565A..94B, 2014ApJS..213...13H, 2014ApJ...792..116Z}. Based on MHD simulations, \citet[][]{2006ApJ...637L.105M} find that the alignment degree between outflow and magnetic field is directly correlated to the magnetic field strength: the stronger the magnetic field, the better the alignment.In the following, we examine the relative position of the outflow axes and the magnetic field directions for CB34, using our near-IR data and compare our result to the findings in the optical and sub-mm wavelength range by \cite{CB34optic}.

In discussing the relative orientation between the CO outflow and the magnetic field, one has to consider that only one component of the spatial orientation of the outflow ($v \, \rm{sin} \, i$) is known from velocity measurements, so projectional effects have to be considered, as well as that polarization segments only trace the magnetic field structure projected on the plane of sky. We assume that the magnetic field is oriented parallel to the polarization segments in the near-IR wavelength range, and perpendicular to the polarization segments in the sub-mm wavelength range. 

The orientation of the CO outflow of CB34 is roughly parallel to the axis linking the cores C1 and C2 \citep[][]{2002A&A...383..502K}, indicating a North-East to South-West orientation.

The orientation of polarization segments in the near-IR wavelength range, tracing the less-dense outer parts of the globule, is almost perpendicular to the direction of the outflow. This fits well to the findings in the optical \citep{CB34optic}. However, the polarization segments in the sub-mm polarization range, tracing the dense center of the globule, are roughly aligned to the outflow direction, especially for those related to core C1 \citep{CB34optic}. 

Our new data is consistent with the finding of a correlation between the magnetic field orientation and the direction of the outflow, depending on the density and distance to the protostellar core.

\subsection{Magnetic fields in star formation as a function of mass}

Here, we are presenting a sample of the lowest mass star-forming clouds that are known, located completely isolated. These low-mass clouds appear to be dominated by magnetic fields \citep[see also, e.g.,][]{2011AJ....142...33A, 2014A&A...565A..94B}.
In larger but still relatively small clouds, such as Taurus, the cloud structure can be explained at least proximately in terms of turbulent simulations \citep[e.g.,][]{2001ApJ...546..980O, 2008AJ....136..404O, 2013MNRAS.436.3247K}. These intermediate mass clouds have comparatively short lifetimes of $\sim 3\,$Myr \citep[e.g.,][]{2001ApJ...562..852H, 2007RMxAA..43..123B,  2016ARA&A..54..135H}. 
In more massive clouds, e.g., Orion (mass $\sim 10^5$M$_{\sun}$), recent work seems to indicate that clouds may have an order of magnitude longer lifetimes \citep[$\sim 30\,$Myr; e.g.,][]{2011ApJ...729..133M, 2015ApJ...806...72M}, and be subject to the influence of “dynamically relevant” magnetic fields \citep[e.g.,][]{2011ApJ...742L...9C, 2013ApJ...762..120P, 2016A&A...590A...2S, 2016ApJ...819..139B}.
In this picture, in order to reconcile the above results including globules, we postulate a mass dependent role of magnetic fields in clouds, whereby magnetic fields appear to dominate at low and high masses, but perhaps are rather sub-dominant at intermediate (e.g., Taurus) masses. 
Why do the globules present magnetic field dominated conditions? The answer to this question, in the absence of more rigorous age or timescale estimates,  requires more detailed modeling of the origin of globules. However, it is consistent with Fig. 1 of \cite{2016A&A...590A...2S} which strongly suggests that at least in the case of B$\,$68 it originated as part of an undulating filamentary structure which then fragmented into cores such as B$\,$68. This can only be explained by magnetic fields (and not gravity, due to the undulation).
The fact that a reasonable fraction of such globules present “filamentary tails” \citep[e.g.,][]{2013A&A...551A..98L} in sub-mm data is consistent with this scenario. The fact that not all globules present such “vestigial tails” may also indicate that globules have comparatively long lifetimes, which would then be consistent with both the fact that they are observed in the starless phases (indicating slow evolution that would then allow for the dissipation of the tails below detectable levels in the outer less dense regions), and the fact that the magnetic fields are observed to be dominant. This tentative scenario must be tested, but it should be noted that proposed reasons for why magnetic fields maybe dominant in high mass clouds are different than for globules.  For example, \cite{2016A&A...590A...2S} propose that massive clouds evolve under their own gravity, inevitably then concentrating the magnetic fields in an initially turbulent cloud medium (see e.g., the differences between the ISF and L1641 within Orion, filaments which both have the same large scale gravitational potential but different inner gravity and density profiles).  In any case, this ansatz should be tested, but for the time being is consistent with observations.  Irrespective of the details of internal evolution of individual systems, the role of magnetic fields in clouds spanning $\sim$5 to 6 orders of magnitude in mass has not been previously discussed, and thus must be tested with future theoretical and observational work.

\section{Conclusions}\label{conc}
For the first time, we have obtained near-IR polarization data and, in conjunction with archival optical and sub-mm data, compiled multi-wavelength polarization maps which allow for the verification of the magnetic field influence on scales of $10^3-10^6$ ($\sim 0.005-5\textrm{pc}$) in the Bok globules CB34, CB56, and [OMK2002]18, covering optically thin and optically thick regions. The major results are:

\begin{enumerate}
\item Our findings determine the distance of CB56 to $\le 200\,$pc, for the first time,  by applying a comparison with measurements of the interstellar polarization $200\,$pc  (see Section~\ref{desc}).
\item We find a strong polarization signal of several percents in all three Bok globules in the near-IR. 
\item In  CB34, the polarization segments in the near-IR  are well aligned with those obtained in the optical tracing scales of $10^5-10^6\,$au ($\sim 0.5-5\textrm{pc}$) as well as with those obtained in the sub-mm which trace the innermost, densest part of the globule on scales of $10^3$ au ($\sim 0.005\textrm{pc}$).
\item In CB56, the polarization segments in the near-IR are well-aligned with those obtained in the optical wavelength range on scales of $10^4-10^5$ au ($\sim 0.05-0.5\textrm{pc}$).
\item In [OMK2002]18, the polarization segments in the western region are well-ordered and show high polarization degrees, while we find indications for depolarizing effects in the eastern region.
\item We find a correlation between the magnetic field structure and the CO outflow of the Bok globule CB34, depending on the density and distance to the protostellar core.
\item For CB34, comparable magnetic field strengths are found in the regions traced in the near-IR resp. sub-mm observations by applying the CF method.
\item We find strong indications for dominant magnetic fields in all of these three globules.
\end{enumerate}

\begin{appendix}\label{appendix}
\begin{figure}[tb]
\includegraphics[width=0.5\textwidth]{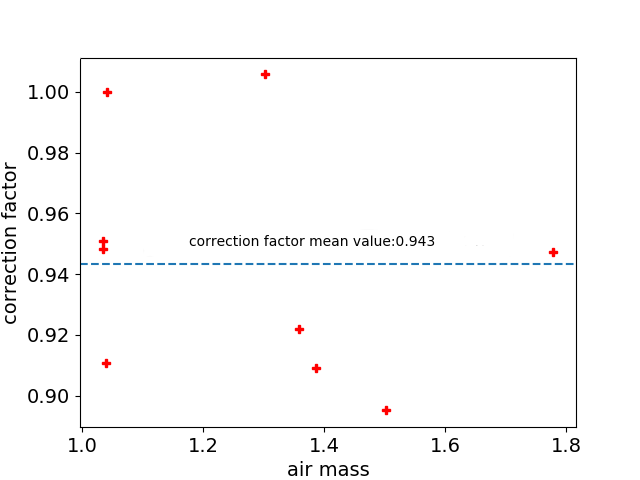}
    \caption{Comparison between the correction factor that minimizes the instrumental polarization and the air mass of the observed unpolarized standard stars. The dashed line correspond to the mean value of the computed correction factors. In contrast to ISAAC/VLT  \citep{2014A&A...565A..94B}, we do not find a correlation between the instrumental polarization and the elevation angle of the telescope, respectively the airmass of the object.}
    \label{correction}
\end{figure}

\section{Correction factor derivation for Wollaston prism}
As stated in Section \ref{data}, a correction factor, $C_\lambda$ for the Wollaston prism has to be calculated to correct for the instrumental polarization, as the transmission ratio of the prism is not ideal. This correction factor is applied to the intensities of the upper, $i_u$, and lower, $i_l$, beams created by the Wollaston prism in the following way \citep{HowToPol}:
\begin{equation}
\begin{aligned}
& i_{u,2} = i_{u,2}*C_\lambda \\
& i_{l,2} = i_{l,1}
\end{aligned}
\end{equation}
In our previous study, we find that the instrumental polarization for an instrument in Nasmyth focus has a strong dependency of the elevation angle of the telescope \citep[ISAAC/VLT; ][]{2014A&A...565A..94B}.  For the here presented study, we test SOFI/NTT for this dependency as well. To determine the instrumental polarization, $C_\lambda$ has to be adjusted in such a way that the measured polarization of the unpolarized standard star is minimal.

From November 11 to November 15, 2015, we observed unpolarized standard stars (see Tab.~\ref{table:standards_summary}) and derive $C_\lambda$ dependent on the elevation angle of the telescope, resp. the airmass of the object (see Fig.~\ref{correction}). We find no correlation between the correction factor, $C_\lambda$, and the elevation angle of SOFI/NTT.  Hence, we use the mean value of the correction factor to reduce our data. Please note that we observed several standard stars with SOFI/NTT over only a few nights. Observations of the same target over a longer period, as performed in \citet{2014A&A...565A..94B}, might result in a different finding.

\begin{acknowledgements}
     The authors thank George A. Gontcharov for providing the data on interstellar polarization in advance to publication. We thank Amelia M. Stutz for valuable discussions which improved this paper. We thank the anonymous referee for a critical and helpful report. GHMB acknowledges financial support from CONICYT through FONDECYT grant 3170657. This project has also received funding from the European Research Council (ERC) under the European Union's Horizon 2020 research and innovation programme (grant agreement No. 757957).
\end{acknowledgements}
\end{appendix}
\bibliographystyle{aa}
\bibliography{references.bib}

\end{document}